\documentstyle[prl,twocolumn,aps,epsf]{revtex}
\begin{document}
\draft

\twocolumn[\hsize\textwidth\columnwidth\hsize\csname
@twocolumnfalse\endcsname

\title{\bf  Systematic   study  of  the   two  band/two  gap
superconductivity \\in carbon-substituted MgB$_2$  by
point-contact spectroscopy}
\author{ Z.  Ho\v lanov\'a,$^{1}$  P. Szab\'o,$^{1}$
P. Samuely,$^{1}$
R. H. T. Wilke,$^{2}$ S. L. Bud'ko,$^{2}$ P. C. Canfield,$^{2}$}
\address{$^1$Centre  of  Low   Temperature  Physics  of  the
Institute of Experimental Physics SAS \& Faculty of Science UPJ\v S,
SK-04353~Ko\v{s}ice,         Slovakia\\
$^2$ Ames Laboratory and Department of Physics and Astronomy,
Iowa State University, Ames, IA 50011 USA}
\date{\today}
\maketitle

\begin{abstract}

Point-contact   measurements   on   the   carbon-substituted
Mg(B$_{1-x}$C$_x$)$_2$   filament/powder   samples  directly
reveal a retention of the two superconducting energy gaps in the
whole doping range from $x  = 0$ to $x \approx 0.1$. The large gap
on the  $\sigma$-band is decreased in an essentially linear
fashion with increasing the carbon concentrations.   The changes
in the the small gap $\Delta_{\pi}$  up to 3.8  \% C are
proportionally smaller and are more difficult to detect but for
the heavily doped sample with $x \approx 0.1$ and $T_c  = 22$ K
both gaps are still present, and significantly reduced, consistent
with a strong essentially linear, reduction  of each gap with the
transition temperature.

\end{abstract}
\date{\today}
\pacs{74.50.+r, 74.70.Ad, 74.62.Dh}

]

The   MgB$_2$   superconductor   \cite{akimitsu}  represents a
challenge for both  technological relevance and fundamental
science. Its relatively high  transition temperature is due to the
interference effect  between the  $\sigma$ and $\pi$ bands with
respectively strongly  and weakly coupled Cooper pairs, and
respectively   large   and   small superconducting energy  gaps.
This theoretical  two-gap   scenario \cite{liu,choi}  has now  been
supported by many experiments
\cite{bouquet,wang,szabo,giubileo,iavarone,suderow,eskildsen,kwok}.
Prospects for applications of MgB$_2$ depend on a success in
increasing  the critical  parameters such as  the upper  critical
field and  critical current. For  that the material  must be
driven from the  clean superconducting  limit to  the dirty one.
The possibility of selective  tuning the $\sigma$ and $\pi$ {\it
intraband} scatterings seems  to be the way of accomplishing this
\cite{wilke,gurevich}. This  can be realized, for  example, via
the substitution of   a constituent element,   such as boron by
carbon. An undesirable side effect may be a  significant
increase of the {\it interband} scattering, presumably  leading
to merging of the two gaps and,  as consequence,  to a substantial
decrease of the transition temperature \cite{liu}.    Previous
studies \cite{ribeiro,samuely1,schmidt1} on  the heavily doped
MgB$_2$ by 10  \% of carbon ($T_c=22$ K) have shown a presence of  the small
energy gap with $2\Delta /k_BT_c \approx$ 1.6, the same  relative
strength as for the gap in  the $\pi$-band of  the undoped system,
incompatible with  the single-gap  scenario. $\mu  ^+$SR experiments have
been performed on the carbon doped MgB$_2$ with $T_c$'s from
38.3 K down to 34.8 K \cite{papagelis}.  The deduced gaps $\Delta _{\pi}$ and
$\Delta _{\sigma}$ showed a  linear decrease with increasing
the carbon content and $T_c$,  but for the samples with $T_c
\le  $   36.1  K  the   reduced  value  of   the  large  gap
2$\Delta/k_BT_c$ was below the BCS weak-coupling value 3.52.

Here   we   present   a   systematic   study   of  the  two-gap
superconductivity         in                 carbon-doped
Mg(B$_{1-x}$C$_x$)$_2$  filaments  with  the  carbon content
$x=$0,  0.021 and  0.038 and  additional experiments  on the
powder sample with $x \approx  $ 0.1. The Andreev reflection
spectra  on the  filaments  show  two very  well  resolved
superconducting energy  gaps. The large  gap on the  $\sigma
$-band   is decreased with   increasing    the   carbon
concentrations, whereas the changes in the smaller gap are
proportionally smaller and, for these low carbon concentrations,
difficult to resolve. For the more heavily doped sample with $x \approx
0.1$ and $T_c = 22$ K, the broadening of the Andreev reflection
spectra makes a direct detection of the large gap difficult  but
we succeeded in detecting it  by  careful measurement at 1.6 K.
Our identification of the large gap was confirmed by the use of an
in-magnetic field experiment where the small gap contribution can
be partially suppressed and the large gap revealed. The large gap
can also be extracted from the analysis of the specific-heat data
\cite{ribeiro}.  In  the 10 \% C doped samples both gaps are
detected and both gaps scale linearly with the suppression of
$T_c$. The temperature dependencies of the large and small gaps
indicate that the interband coupling is not significantly changed
upon the doping in the whole concentration range.

Wire segments of carbon-substituted MgB$_2$ were synthesized from
carbon doped  boron filaments  made by  chemical vapor deposition.
The  boron and carbon were  co-deposited to form a fiber of about
75 $ \mu$m diameter. Mg(B$_{1-x}$C$_x$)$_2$ filaments were made by
exposing the fibers to Mg vapor.  More details about preparation
are given elsewhere \cite{wilke}. X-ray diffraction shows the
MgB$_2$ phase and Mg lines. No traces of MgO and B$_4$C have  been
found. No measurable changes in the $c$-axis lattice parameter
were detected, whereas the measured changes  of the $a$-axis
lattice parameter $\Delta a(x) $ were used to determine the carbon
concentration by comparison to Avdeev {\it et al.} neutron
difraction  \cite{avdeev} on $x \sim 0.1$ samples of
Mg(B$_{1-x}$C$_x$)$_2$. The resulting carbon concentrations are
$x=$0, 0.021 and 0.038. The quality of the samples is evidenced by
a single step and narrow transitions to superconducting state in
magnetization and resistivity with $T_c = $ 39, 37.5 and 36.2 K
defined at the onset of  the superconducting state with  $\delta
T_c \le$  0.3 K between  10 and  90  \%  of the normal state
resistivity. The resistivity  of the  wires increases  from 0.5
$\mu \Omega$cm for the undoped system to about 10 $\mu \Omega$cm
for 3.5 \%  C. Mg(B$_{0.9}$C$_{0.1})_2$ with $T_c$ = 22 K  in the
form  of pellets were  prepared following the procedure  described
in Ref.\cite{ribeiro}  from magnesium lumps and  B$_4$C powder.
Traces of  B$_4$C were not visible in the  XRD patterns. Small
amounts of two  impurity phases (MgO  and MgB$_2$C$_2$)  resulted
even  with optimization of the synthesis.

Point-contact  measurements on  the wire  segments have been
performed  in a  "reversed" configuration  - with  the
Mg(B$_{1-x}$C$_x$)$_2$ wire as a tip touching softly a bulk piece
of electrochemically  cleaned copper.  In   the  experiments  on
bulk Mg(B$_{0.9}$C$_{0.1})_2$      samples     a      tip     was
(electro)chemically formed  from copper, platinum  or silver
wires.  A  special  point-contact  approaching  system  with a
negligible   thermal  expansion   allows  for  temperature
dependent  measurements  up  to  100  K.  A standard lock-in
technique  at 10 kHz was  used to  measure the differential
resistance  as a  function of  applied voltage  on the point
contacts.  The  microconstrictions  were  prepared  {\it  in
situ}. The  approaching system enabled both  the lateral and
vertical  movements   of  the  tip   by  differential  screw
mechanism.

The point-contact  spectrum - {\it  differential conductance
versus voltage} between a metal  and a superconductor can be
compared with the Blonder, Tinkham and Klapwijk (BTK) theory using
as  input  parameters  the  energy  gap $\Delta$, the parameter
$z$ (measure  for the  strength of  the interface barrier with
transmission coefficient T$  = 1/(1+z^2)$) and a parameter
$\Gamma$  for   the  quasi-particle   lifetime broadening
\cite{plecenik}.  In  the  case  of  the MgB$_2$ two-gap
superconductor  the   overall  conductance  can  be expressed as a
weighted sum  of the partial BTK conductances from the quasi
two-dimensional  $\sigma$-bands (with a large gap $\Delta
_{\sigma}$) and the 3D $\pi$-bands (with a small gap  $\Delta
_{\pi}$)  $\Sigma   =  \alpha  \Sigma_{\pi}  + (1-\alpha
)\Sigma_{\sigma}$. The weight factor $\alpha $ for the $\Delta
_{\pi} $ gap contribution can vary from 0.65 for the
tunneling/point-contact current  strictly in the MgB$_2$
$ab$-plane  to 0.99  of $c$-axis  tunneling \cite{brinkman}.
Indeed,  the  tunneling  experiments  on  single crystalline
MgB$_2$ \cite{eskildsen}  have proved that  the small gap  on the
$\pi$-band is observed for any tunneling direction while $\Delta
_{\sigma}$  is   observable  only   for  significant $ab$-plane
tunneling component.

Extensive measurements  on tens of  pieces of wire  segments and
bulk  doped MgB$_2$ have  been performed. The  point-contacts revealed
different barrier transparencies from more metallic interface with
$z = 0.4 $ up to an intermediate case between metallic and
tunneling barrier with $z \sim $ 1.2. Crystallites in the wire
segments have  a size up to tens of microns.  For such large
crystallites, the point-contacts yield  an information on a
particular single crystal at the junction,  but of unknown
orientation. By trial and error  we looked for the junctions where
the both gaps would be present. A large number of the "two-gap"
spectra on Mg(B$_{1-x}$C$_x$)$_2$ wires of $x = $ 0, 0.021 and
0.038  carbon concentrations  have been  recorded. For the
presentation and analysis we have chosen those with a small
broadening $\Gamma _i $, typically  less than 10 per cent of
respective gap $\Delta _i$. In accordance with the previous report
\cite{samuely1}    the most of the  spectra on
Mg(B$_{0.9}$C$_{0.1})_2$ revealed apparently  only the small
energy  gap   $\Delta  _{\pi}$.  The broadening  parameter $\Gamma
$ was in  this case at least  20 per cent of  the gap value.
Larger
scattering  rates  and inhomogeneities in the heavily doped
Mg(B$_{0.9}$C$_{0.1})_2$   are  probably   the  main  reason
preventing as easy of a
direct detection of the large gap, here. Since the point-contact
spectroscopy is  a surface sensitive technique it is sensitive  to
a  possible  surface  proximity effect with correspondingly
suppressed  $T_c$. For most of the junctions the particular  $T_c$  has
been   checked:  in  all cases  the point-contact  transition
temperature  agreed with the bulk $T_c$. In addition, as it will
be shown below, both gaps close near the bulk $T_c$.

Figure  1 shows  characteristic examples  of the  normalized
conductance-versus-voltage spectra (full  lines) obtained on the
Mg(B$_{1-x}$C$_{x})_2$-metal   junctions.    All   displayed
point-contact  conductances  have  been  normalized  to  the
conductance in  the normal state  at $ T  > T_c$. The fits to the
spectra by the two-gap BTK formula are shown by open circles. As
can be seen the spectra  on the  wires with  the lower
carbon-concentrations - up   to  3.8   \%  C   reveal  very   well
resolved   two superconducting  energy  gaps  with  small  {\it
intrinsic} broadening. The upper spectrum was obtained on the more
heavily doped, 10 \% C, powder sample. To achieve better
resolution this spectrum is measured at 1.6 K, whereas  the others
were measured at 4.2  K. A one-gap fit to this spectrum is shown
as well (by the open triangles). Obviously, the two-gap formula fits the spectrum better.
Thus, the spectrum represents one  of a few examples indicating a
presence of the large gap in MgB$_2$ with 10 per cent of carbon.

Whereas we are able to detect the larger gap in the 1.6 K
tunnelling spectra of the Mg(B$_{0.9}$C$_{0.1}$)$_2$ sample, it is
not as unambiguously resolved as in the lower doped samples. Our
identification of the larger gap is strongly supported by in-field
tunneling measurements. In the Fig. 2 the spectrum on the 10 \% C
MgB$_2$ sample is shown in magnetic field. In the spectrum
recorded at zero magnetic field the shoulder originating from the
second large gap is not very pronounced  and we have to rely on
the fit showing that the two gap formula works better than the
one-gap fit. But a finite magnetic field partially suppresses the
contribution from the $\pi$-band, with  a smaller gap, and makes a
presence  of the large  gap  from the $\sigma$-band more
conspicuous. This is a well known  effect in MgB$_2$  which we
detected  in our original paper on the two-gap superconductivity
in pure MgB$_2$ \cite{szabo}. Here, in a  very similar way it
helps  to  further  reveal  the  existence  of  the large gap
$\Delta_{\sigma}$:
appearing as  more and more clearly defined shoulder, best seen in
the spectrum taken at $B=$ 0.7 Tesla. At even higher fields the
intensity of the $\pi$-band contribution  is suppressed so that
the $\pi$ and $\sigma$ peaks interfere to the single maximum,
which   is located indeed   at higher  voltages \cite{samuely2}
than the original maximum of  the prevailing $\pi$-band spectrum
at $B=0$. Data such as these are further evidence of the clear
persistence of two superconducting gaps in Mg(B$_{1-x}$C$_x)_2$
even at $x =$ 0.1 levels.

To  obtain further  information on  the large  energy gap $\Delta
_{\sigma}$ in the heavily doped Mg(B$_{0.9}$C$_{0.1})_2$, an
analysis of  the electronic specific  heat on the  data from
Ribeiro {\it et al.} \cite{ribeiro} was performed within the frame
work of the $\alpha$-model as proposed  for MgB$_2$ by Bouquet
{\it  et al.}  \cite{bouquet}.   A typical jump at $T_c$  from the
large energy gap was  seen as well  as an excess  weight at low
temperatures  as a  hint of the small energy  gap $\Delta_{\pi}$.
Even, if  this is an indirect method the resulting gaps are
indicated as asterisks in  Fig. 3 and are in good agreement with
the   direct   point-contact determination of the gaps.

Figure 3 shows the overall statistics  of energy gaps which have
been obtained from  analysis done on about 15 best resolved spectra
for each  carbon concentration, except for
the 10  \% C substituted  MgB$_2$ sample, where  we only two
junctions revealed the large gap. The widths of the bars
indicates  the error  in determination  of the  energy gap
associated with the fitting  to  the  BTK   model.  The double
peak distribution of the large and small gaps well covers the gap
sizes given by most of  the tunneling and point contact data on
the pure MgB$_2$ \cite{szabo,iavarone,suderow,eskildsen,schmidt2}.
Due to distribution of the two gaps better comparison can be
made with the similar histograms
obtained from the MgB$_2$  point-contact data by Naidyuk {\it et
al.} \cite{naidyuk} and Bugoslavsky {\it et al.}
\cite{bugoslavsky}.  Good  agreement  is  found  among those
measurements. There is also a possibility to make comparison
 with the distribution  of the gaps on the
Fermi  surface calculated  by Choi  {\it et  al.} within the fully
anisotropic Eliashberg formalism \cite{choi}. Matrinez-Samper {\it
et  al.} \cite{suderow} even fitted directly their  particular
tunneling  spectra with  a distribution  of gaps showing two maxima
around 2.5 and 7 meV. But recently  Mazin {\it  et al.}  argued
that to see directly more than  two gaps  in the  spectrum or even
whole distribution would  require  extremely   small, unrealistic
$\pi$  and $\sigma$  {\it  intraband}  scattering rates
\cite{mazin}. Thus, this particular question of the gap
distribution around two maxima asks for further studies.
Nevertheless, even with the presented gap distributions  one can
see clearly the effect of the carbon doping on the gaps.

In Fig. 4 the gap energies as a function of $T_c$ are presented.
The circles  are positioned at the  averaged energies of the
gap distributions in Fig.3 and  the error bars represent the
standard deviations of the distributions.
Broadly, the tendency of the evolution of two gaps in MgB$_2$ with
carbon doping seems to be almost linear with both gaps disappearing (as
extrapolated) simultaneously at $T_c =$ 0. More data is needed to
address possible non-linearity at low carbon content and to more
precisely determine the point where two gaps merge to one or
disappear (the error bars on the data presented in Fig. 4 support
a merging of the two gaps for 0 K $\le T_c \le$ 10 K). The inset
to Fig. 4 with the reduced gaps 2$\Delta/k_BT_c$,  shows that
already for the small carbon concentrations  there  is  a tendency
to merge both gaps. Despite this  tendency, on the samples with 10
per cent of carbon doping and $T_c=$ 22K  the two-gap
superconductivity is still clearly retained as  presented by  the
normalized gaps 2$\Delta_{\sigma}/k_BT_c     \simeq 3.6$ and
2$\Delta_{\pi}/k_BT_c \simeq 1.6$.

The  temperature dependence  of the  large and  small energy gaps
$\Delta  _{\sigma}$  and   $\Delta  _{\pi}$  from  the
point-contact spectra on  Mg(B$_{1-x}$C$_x$)$_2$ samples with
different carbon concentrations  are shown in  Fig. 5. One  can
notice that the shape of the temperature dependence of both gaps
does not  show  any  obvious  change  due  to  a different carbon
concentration. As  mentioned already by  Suhl {\it et  al.}
\cite{suhl}, the shape of  the temperature dependence  of the
small gap is related to the measure of  the interband coupling. If
it was too small the  small  energy gap  would   tend  to  close  at
temperatures much below  the  bulk  $T_c$  with  a tail to $T_c$.
In the case of pure MgB$2$ the temperature dependence  of the
small gap slightly undershoots  the  BCS  line  \cite{szabo}  in
line with the predictions of Liu {\it et al.} \cite{liu}. Within
the shown error bars in  Fig. 5  we can  conclude that  the
interband coupling is not changed by the carbon doping.

The theoretical  calculations have shown that  due to a very
different $k$-space  distribution of the  $\pi$ and $\sigma$
bands,  the  only  route  to  increase  the  $\sigma  - \pi$
scattering is  via interlayer hopping, from  a $p_z$ orbital
($\pi$-band)  in   one  atomic  layer  to   a  bond  orbital
($\sigma$-band)  in another  layer \cite{mazin2}.  But  no changes
in the $c$-lattice parameter which could help the interlayer
charge transfer  in carbon  substituted samples  in comparison
with the   pure  MgB$_2$   were  detected   \cite{wilke}.  Recent
theoretical modelling  also corroborates that  carbon doping is
not  favorable   for  increased   interband  scattering
\cite{erwin}.  What is  then  the  reason for  a significant
decrease of  the transition temperature  to $T_c =$  22 K at 10 \%
C doping? The  transition temperature is determined by the
characteristic phonon  frequency, the  strentgh of  the
electron-electron coupling and the density of the conduction
electrons. Due to hole  $\sigma$-band filling the density of
states can be decreased  by carbon substituting boron. Masui {\it
et  al.}  have  detected  hardening  and narrowing the $E_{2g}$
phonon modes by Raman scattering \cite{masui} which can  have a
dramatic effect  on the  strength of the cooper pairs coupling in
the $\sigma$-band. Decrease in the density of states at the Fermi
level was suggested by band structure calculations
\cite{medvedeva} and observed through measurements of heat
capacity \cite{ribeiro}. Also a significantly lower anisotropy in
$H_{c2}$ \cite{ribeiro} in heavily carbon doped compound implies
that  the $\sigma$-band Fermi surface  is not nearly so 2D as in
the pure MgB$_2$.

In conclusion, we have obtained an experimental evidence for
existence   of   the   two-gap   superconductivity   in  the
carbon-substituted MgB$_2$  for all C  concentrations from 0
to about 10 per cent with $T_c's$  from 39 down to 22 K. The
both gaps are closing near  the bulk critical temperature of
the  respective sample.  The temperature  dependence of  the
gaps indicates  no changes in  the {\it interband}  coupling
which could eventually lead to definite merging of the gaps.


This  work  has  been  supported  by  the Slovak Science and
Technology     Assistance      Agency     under     contract
No.APVT-51-020102.  Centre  of  Low  Temperature  Physics is
operated as  the Centre of Excellence  of the Slovak Academy
of Sciences under contract  no. I/2/2003. Ames Laboratory is
operated  for the  U.S. Department  of Energy  by Iowa State
University under  Contract No. W-7405-Eng-82.  This work was
supported  by the  Director for  Energy Research,  Office of
Basic   Energy  Sciences.   The  liquid   nitrogen  for  the
experiment has  been sponsored by the  U.S. Steel Ko\v sice,
s.r.o.

\pagebreak
\onecolumn


\begin{figure}
\epsfverbosetrue
\hspace{350mm}
\epsfxsize=10cm
\epsfysize=12cm
\begin{center}
\hspace{550mm}
\epsffile{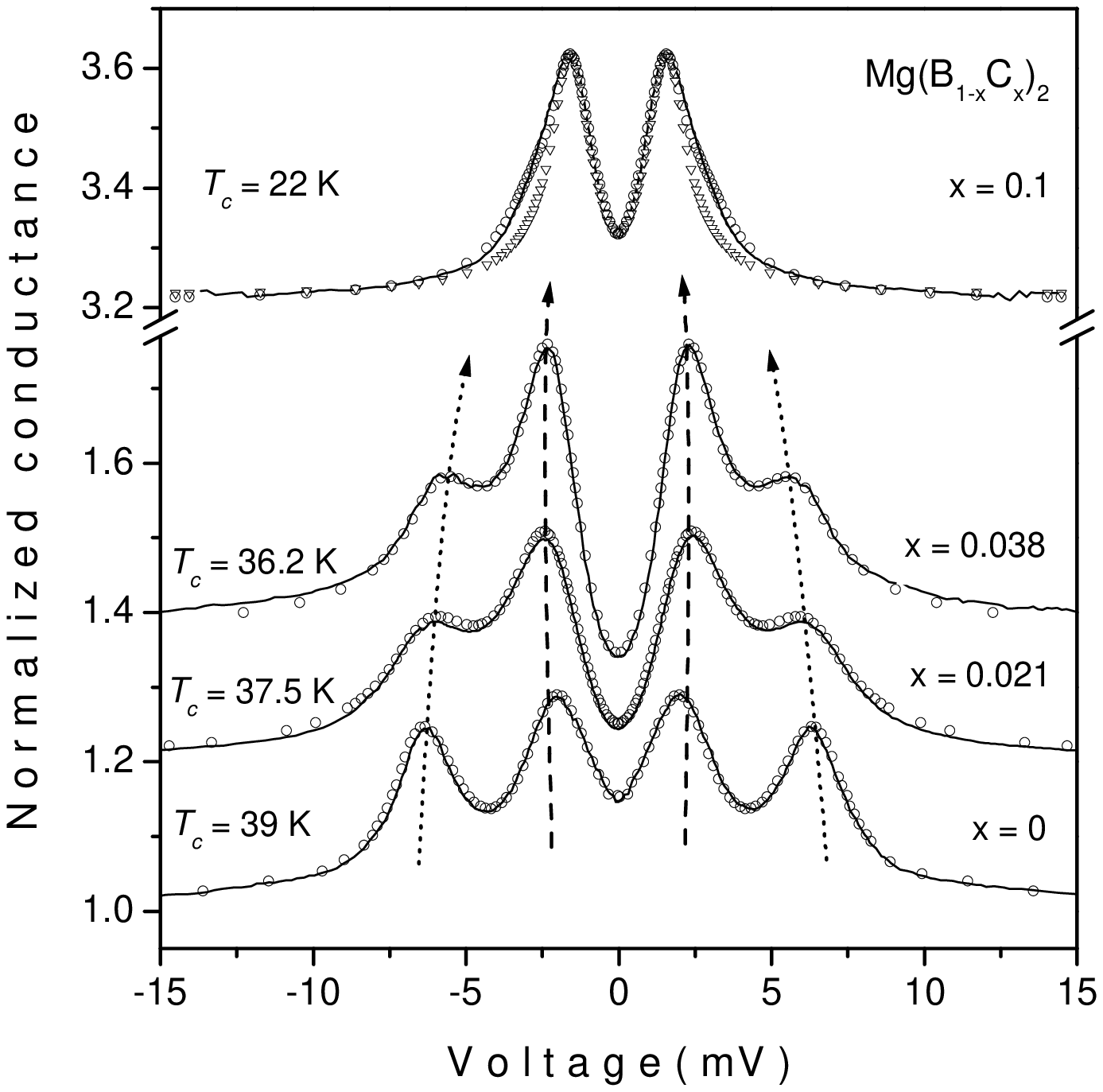}
\vspace{2mm}
\end{center}
\caption{
Full lines - Mg(B$_{1-x}$C$_{x}$)$_2$  point-contact spectra  at $T
=$ 4.2  K (except for the  top spectrum recorded at  1.6 K).
Open circles - fitting
for  the  thermally  smeared  BTK  model  for two gaps. Open
triangles for $x=0.1$ curve
 - fit to the one-gap BTK formula.  The upper curves are vertically
shifted for  the clarity. The arrows  indicate a development
of the gaps with increasing carbon content.
}
\end{figure}

\newpage

\begin{figure}
\epsfverbosetrue
\hspace{350mm}
\epsfxsize=10cm
\epsfysize=12cm
\begin{center}
\hspace{550mm}
\epsffile{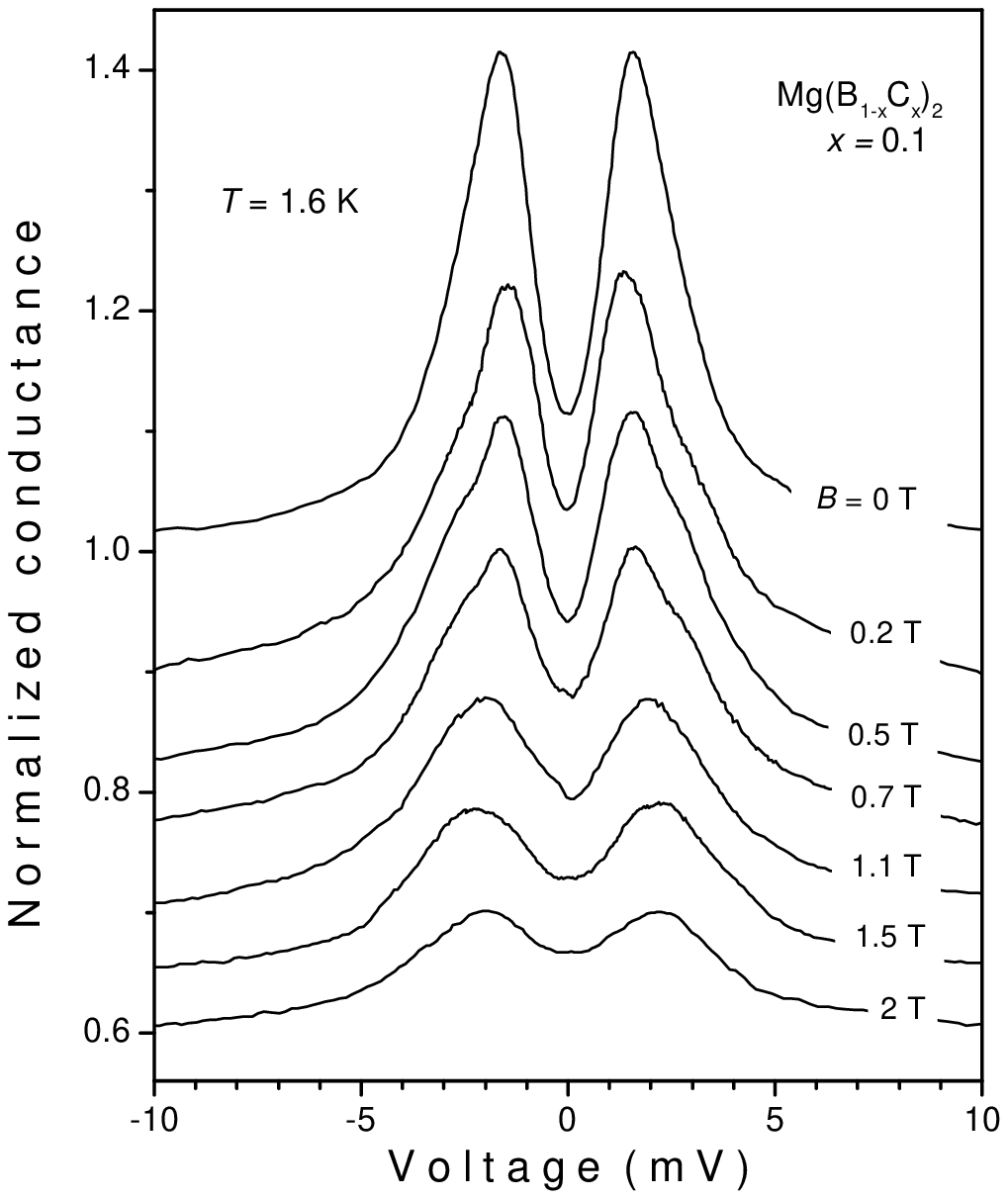}
\vspace{2mm}
\end{center}
\caption{Mg(B$_{0.9}$C$_{0.1}$)$_2$-Cu         point-contact
spectrum
at  1.6.K (shown  in Fig.1)  recorded at  different magnetic
fields.  Partial suppression  of the  small gap contribution
helps in  revealing  the large gap (see the text). The curves
at finite field are shifted down for clarity.
}
\end{figure}

\newpage

\begin{figure}
\epsfverbosetrue
\hspace{350mm}
\epsfxsize=10cm
\epsfysize=12cm
\begin{center}
\hspace{550mm}
\epsffile{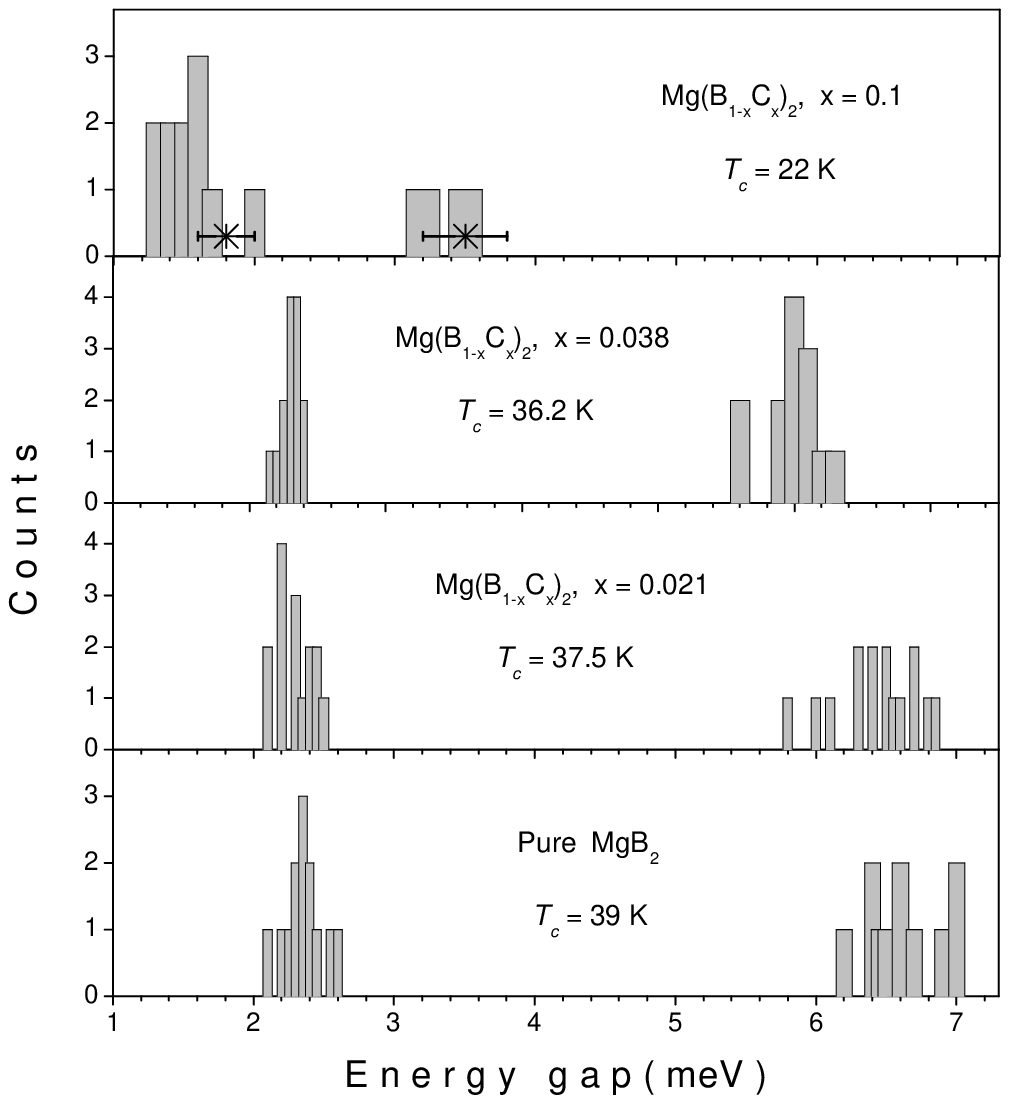}
\vspace{2mm}
\end{center}
\caption{Distribution  of the  superconducting  energy  gaps  of
Mg(B$_{1-x}$C$_{x}$)$_2$    as   obtained    from   numerous
point-contact   junctions  on   different  samples   of  the
particular C concentration.
Asterisks -  the  gaps obtained  from  analysis  of the  specific heat
[14].}
\end{figure}

\newpage

\begin{figure}
\epsfverbosetrue
\hspace{350mm}
\epsfxsize=11cm
\epsfysize=10cm
\begin{center}
\hspace{550mm}
\epsffile{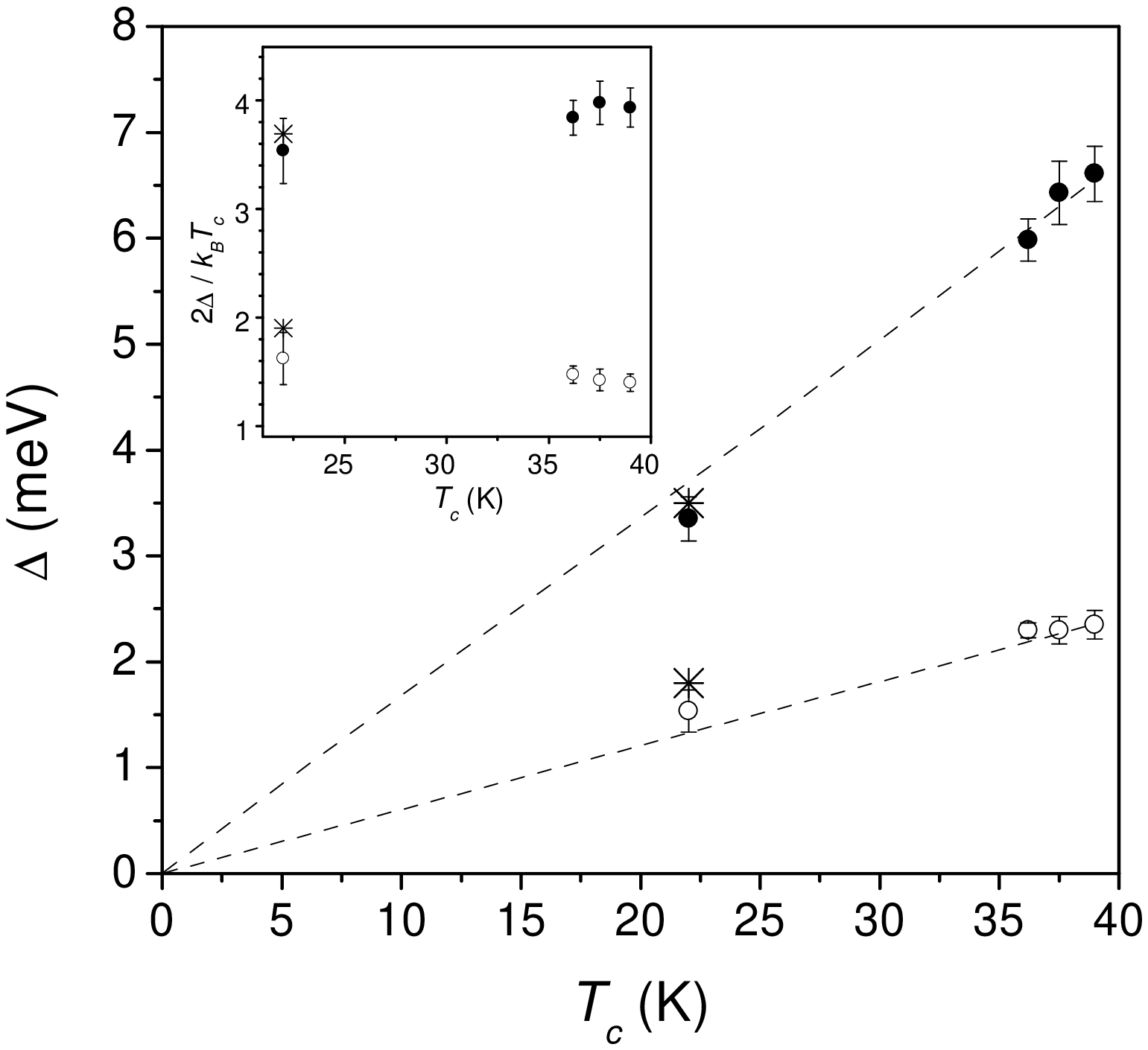}
\vspace{2mm}
\end{center}
\caption{Superconducting  energy   gaps  from  point-contact
experiments at lowest temperature as
a function  of  $T_c$. The points   -  average  energy  of  the
particular gap distribution shown in  Fig.3. The error bars -
the  standard  deviations  in  the  distributions. Asterisks
- deduced  from analysis  of the  specific heat  measurement
(see text). Lines are guide for the eye.
}
\end{figure}

\newpage

\begin{figure}
\epsfverbosetrue
\hspace{350mm}
\epsfxsize=10cm
\epsfysize=7cm
\begin{center}
\hspace{550mm}
\epsffile{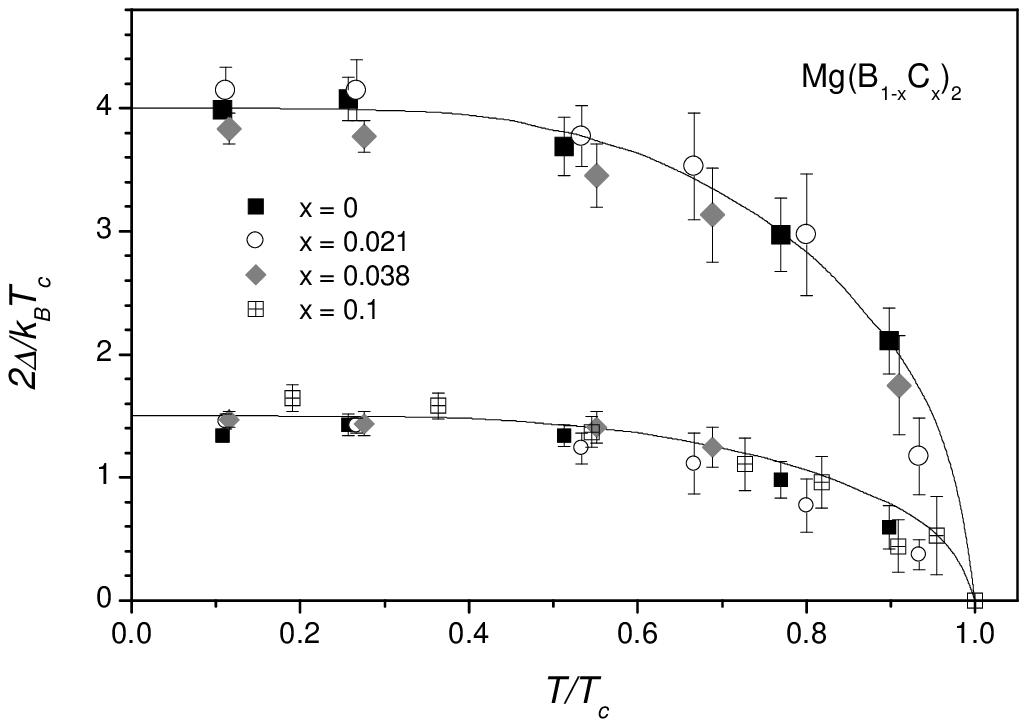}
\vspace{2mm}
\end{center}
\caption{Temperature dependence of the energy gaps
of Mg(B$_{1-x}$C$_{x}$)$_2$ determined
from the fitting of the point-contact spectra.
Full lines represent the BCS  prediction scaled to the $\pi$
and $\sigma$ gaps.}
\end{figure}


\begin{references}
\bibitem{akimitsu} J. Nagamatsu, N.Nakagawa, T. Muranaka, Y.
Zenitani,  J.  Akimitsu,  Nature   (London)  {\bf  410},  63
(2001).
\bibitem{liu} A. Y. Liu, I. I. Mazin, and J. Kortus,
Phys. Rev. Lett. {\bf  87}, 087005 (2001).
\bibitem{choi} H.
J.  Choi et  al., Nature  {\bf 418},  758 (2002); Phys. Rev.
B {\bf  66},  020513  (2002).
\bibitem{bouquet}  F. Bouquet
{\it   et   al.},   Phys. Rev. Lett.   {\bf   87},  047001 (2001).
\bibitem{wang}  Y. Wang
{\it   et   al.},   Physica C   {\bf   355},  179 (2001).
\bibitem{szabo} P.  Szab\'o {\it et  al.}, Phys. Rev.  Lett.
{\bf  87},  137005  (2001).
\bibitem{giubileo}  F. Giubileo
{\it  et al.},  Phys. Rev.  Lett. {\bf  87}, 177008  (2001).
\bibitem{iavarone}  M.  Iavarone  {\it  et  al.}, Phys. Rev.
Lett.   {\bf  89},   187002  (2002).
\bibitem{suderow}  P.
Martinez-Samper  {\it  et  al.},  Physica  C  {\bf385},  233
(2003).
\bibitem{eskildsen} M.  R. Eskildsen  {\it et al.},
Physica C  {\bf385}, 169 (2003).
\bibitem{kwok} Review {\it
Superconductivity   in  MgB$_2$:   Electrons,  Phonons   and
Vortices}, eds. W. Kwok, G. Crabtree, S. L. Bud'ko and P. C.
Canfield,   Physica   C   {\bf   385},   Nos.   1-2  (2003).
\bibitem{wilke}   R.    H.   T.   Wilke    {\it   et   al.},
cond-mat/0312235.
\bibitem{gurevich}  A.  Gurevich  {\it et
al.},   Supercond.  Sci.   Technol.{\bf  17},   278  (2004).
\bibitem{ribeiro} R. A. Ribeiro,  S. L. Bud'ko, C. Petrovic.
and  P.  C.  Canfield,  Physica  C  {\bf  384},  227 (2003).
\bibitem{samuely1} P. Samuely et al., Phys. Rev. B {\bf 68},
020505  (R) (2003).
\bibitem{schmidt1} H.  Schmidt et  al.,
Phys. Rev.  B {\bf 68}, 060508  (R) (2003).
\bibitem{papagelis}
K. Papagelis,  J. Arvanitidis, K. Prassides,  A. Schenck, T.
Takenobu,  and  Y.  Iwasa,  Europhys.  Lett.  {\bf  61}, 254
(2003).
\bibitem{avdeev} M. Avdeev,
J.D.  Jorgensen, R.  A. Ribeiro,   S. L.  Bud'ko, and  P. C.
Canfield.    Physica    C    {\bf    387},    301    (2003).
\bibitem{plecenik}  A.Plecenik  {\it  et  al.},  Phys.  Rev.
B {\bf  49},  10016  (1996).
\bibitem{brinkman} A. Brinkman
{\it  et al.},  Phys. Rev.  B {\bf  65}, 180517  (R) (2002).
\bibitem{samuely2}   P.  Samuely   {\it  et   al.},  Physica
C {\bf385}, 244 (2003).
\bibitem{schmidt2} H. Schmidt, J.F.
Zasadzinski, K.E. Gray and D.G.  Hinks, Physica C {\bf 385},
221 (2003).
\bibitem{naidyuk} Yu. G. Naidyuk  {\it et al.},
Pis'ma Zh. Eksp. Teor. Fiz. {\bf 75}, 283 (2002); JETP Lett.
{\bf 75},  238 (2002).
\bibitem{bugoslavsky}  Y. Bugoslavsky
{\it  et  al.},  cond-mat/0307540.
\bibitem{mazin} I. Mazin
{\it  et  al.},  Phys.  Rev.  B  {\bf  69},  056501  (2004).
\bibitem{suhl}
H. Suhl, B. T. Matthias, L. R. Walker, Phys. Rev. Lett. {\bf 3}, 552 (1959).
\bibitem{mazin2}
I.  I.  Mazin  {\it et al.},  Phys.  Rev.  Lett. {\bf 89}, 107002
(2002).
\bibitem{erwin}   S.    C.   Erwin   and    I.   I.   Mazin,
Phys. Rev. B {\bf 68}, 132505 (2003).
\bibitem{masui}
T. Masui, S. Lee, and S. Tajima, cond-mat/0312458.
\bibitem{medvedeva}   N. I. Medvedeva {\it et al.},
Phys. Rev. B {\bf 64}, 020502 (2001).

\end{references}
\end{document}